\documentclass[twocolumn, aps, prl, superscriptaddress]{revtex4-2}
\usepackage{mathtools}
\usepackage{newtxtext}
\usepackage[T1]{fontenc}
\usepackage{amssymb}  
\usepackage{amsfonts}  
\usepackage{amsmath}  

\usepackage{amsthm}  
\usepackage{bm}  

\usepackage{bbm}  
\usepackage{xcolor}
\usepackage[colorlinks, allcolors=purple]{hyperref}
\usepackage{graphicx}
\usepackage{tabularx}
\usepackage{bbding}
\usepackage{comment}
\usepackage{float}
\usepackage{natbib}
\usepackage{ifthen}
\usepackage{cleveref}
\usepackage{lipsum}
\usepackage{makecell}
\usepackage{braket}

\theoremstyle{definition}

\theoremstyle{remark}

\usepackage{tikz, datatool, xcolor, ifthen, csvsimple}
\usetikzlibrary{arrows, shapes, math, decorations, decorations.markings}
\usetikzlibrary{calc}
\usetikzlibrary{shapes.geometric}
\usetikzlibrary {arrows.meta}
\usepackage{pgfplots}
\pgfplotsset{compat=newest}
\usepackage{pgfmath}
\input pgfmath.tex

\begin{document}
\title{Competing $s$-wave pairing in overdoped $t$-$J$ model}

\begin{abstract}
The $d$-wave pairing symmetry has long been considered a defining feature of high-temperature superconductivity in cuprates.
In this work, we reveal that $s$-wave pairing states exhibit variational energies comparable to the $d$-wave state in a square $t$-$J$ model, particularly at high doping levels ($\delta\gtrsim  15\%$) by using the state-of-the-art tensor network simulation.
This surprising result suggests that $s$-wave pairing may play an important role in the cuprate phase diagram,
especially for the overdoped region. 
Our findings provide a potential resolution to discrepancies in recent Josephson tunneling experiments on twisted bilayer cuprates and offer new insights into the evolution of pairing symmetry with doping. 
\end{abstract}

\author{Wayne Zheng}
\thanks{These authors contribute equally.}
\affiliation{Department of Physics, The Chinese University of Hong Kong, Sha Tin, New Territories, Hong Kong, China}

\author{Tao Cheng}
\thanks{These authors contribute equally.}
\affiliation{State key laboratory of quantum functional materials and Department of Physics, Southern University of Science and Technology, Shenzhen 518055, China}

\author{Zheng-Yuan Yue}
\affiliation{Department of Physics, The Chinese University of Hong Kong, Sha Tin, New Territories, Hong Kong, China}

\author{Fu-Chun Zhang}
\affiliation{Kavli Institute for Theoretical Sciences, University of Chinese Academy of Sciences, Beijing 100190, China 2Collaborative Innovation Center of Advanced Microstructures, Nanjing University , Nanjing 210093, China}

\author{Wei-Qiang Chen}
\email{chenwq@sustech.edu.cn}
\affiliation{State key laboratory of quantum functional materials and Department of Physics, Southern University of Science and Technology, Shenzhen 518055, China}
\affiliation{Quantum Science Center of Guangdong-Hong Kong-Macao Greater Bay Area, Shenzhen 518045, China}

\author{Zheng-Cheng Gu}
\email{zcgu@phy.cuhk.edu.hk}
\affiliation{Department of Physics, The Chinese University of Hong Kong, Sha Tin, New Territories, Hong Kong, China}

\maketitle

\paragraph*{Introduction ---}

The discovery of high-temperature (high-$T_{c}$) cuprate superconductors~\cite{Bednorz1986, RevModPhys.78.17, Keimer2015} decades ago is a major breakthrough in modern condensed matter physics, and it is widely accepted that the pairing mechanism is closely related to the short-range antiferromagnetic (AFM) fluctuations~\cite{science.235.4793.1196}, fundamentally distinguished from the BCS mechanism. 
Theoretically, the Hubbard model and $t$-$J$ model have been considered to be the simplest models that capture the main physics of the cuprate, where many studies have suggested that its ground state is a $d$-wave superconductor~\cite{RevModPhys.78.17}.
Various experimental evidences, from angle-resolved photoemission spectroscopy, scanning tunneling microscopy, to phase--sensitive experiment, have also confirmed a predominant $d$-wave pairing symmetry in cuprates~\cite{PhysRevLett.70.1553,PhysRevLett.71.2134,PhysRevLett.73.593,RevModPhys.72.969}.
However, there are still some debates on the pairing symmetry of the cuprate superconductor.  Some refined experiments in the mid-1990s revealed the possible coexistence of superconducting gaps in $s$- and $d$- waves in cuprates~\cite{PhysRevB.50.6530,science.267.5199.862,Muller1995}.
Subsequent and more deliberate theoretical analysis of AFM spin fluctuations further supports these observations, suggesting a possibly more complex superconducting phase diagram than previously anticipated in cuprates with isotropic and anisotropic pairing gaps~\cite{cm9603046,Koltenbah_1997}.
In particular, Refs.~\cite{cm9603046,Koltenbah_1997} both suggested a $d$-wave at low doping levels while a competing $s$-wave at high doping levels.
Experimentally, recent Josephson junction experiments of twisted cuprates also show controversial results~\cite{PhysRevX.11.031011, Wang2023, science.abl8371}. 

In the weak-coupling limit, heuristic arguments suggest that $d$-wave pairing symmetry is inherently preferred by repulsive interactions. This preference arises from the ability of Cooper pairs with a node at the origin. 
However, this simplistic view assumes that the pairing wavefunction extends extensively in real space, a premise that may not hold in the strong-coupling limit.
Recent tensor network simulations on the Kagome lattice $t$-$J$ model suggest that, as metastable states, extensive $s$-wave pairing can exhibit lower energy than $d+\text{i}d$-wave pairing across a broad range of doping levels\cite{KagometJ}.
Nevertheless, the true ground state might eventually be dominated by charge density wave (CDW) or non-Fermi liquid phases. Furthermore, a separate study on the honeycomb $t$-$J$ model indicates that extensive $s$-wave pairing could represent a highly favorable metastable state, with variational energy comparable to that of the $d+\text{i}d$-wave ground state.

In this study, we present an unexpected discovery indicating that an $s$-wave pairing state could potentially be induced on the bi-layer twisted square lattice $t$-$J$ model at large doping levels. We note that while the $d$-wave pairing state exhibits significantly lower variational energy at lower doping concentrations, it converges to an almost degenerate state with the $s$-wave pairing state when the doping level $\delta \gtrsim 15\%$.
To delve deeper into this phenomenon, we further developed a Ginzburg-Landau model to describe the possible emergence of mixture of d-wave and s-wave pairing states in a realistic twisted bi-layer cuprates model. This model is then applied to elucidate the recent experimental observations of Josephson current in cuprate materials twisted by larger angles.

\begin{figure}[t]
    \centering
    \includegraphics[width=0.45\textwidth]{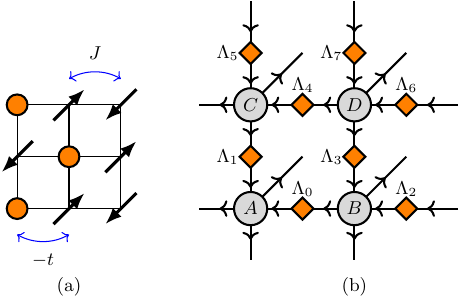}
    \caption{
        (a) Schematic $t$-$J$ model on a square lattice.
        (b) A fTPS comprised of a $2 \times 2$ unit cell on a square lattice.
    }
    \label{fig:tJ_TN}
\end{figure}

\paragraph*{Tensor network variational ansatz for the $t$-$J$ model ---}
The well-accepted one-band $t$-$J$ model on a square lattice is widely believed to be able to capture the essence of high-$T_{c}$ superconductivity~\cite{PhysRevB.37.3759, BASKARAN1993853, RevModPhys.66.763, RevModPhys.78.17}:
\begin{equation}
    \begin{aligned}
    H_{tJ}
    =
    -t \sum_{\langle{ij}\rangle,\sigma}
    \left(
       \hat{c}_{i\sigma}^{\dagger}\hat{c}_{j\sigma}+h.c.
    \right) 
    +J\sum_{\langle{ij}\rangle}\left(\mathbf{S}_{i}\cdot\mathbf{S}_{j}-\frac{n_{i}n_{j}}{4}\right).
    \end{aligned}
    \label{Eq:ham_tJ}
\end{equation}
where $\hat{c}_{j\sigma}$ is the fermion annihilation operator in no-double-occupancy subspace.  $n_{j}\equiv\sum_{\sigma}\hat{c}_{j\sigma}^{\dagger}\hat{c}_{j\sigma}$ is the particle number operator on site $j$.
$\langle{ij}\rangle$ denotes a nearest-neighbor (NN) pair of sites.  
Its ground state phase diagram could be studied by the fermionic tensor product state (fTPS):
\begin{equation}
    \ket{\Psi}
    =\sum_{\{s\}}f\text{-Tr}\left(T^{s_{0}}\Lambda\cdots{T}^{s_{N-1}}\Lambda\right)\vert{s}_{0}\cdots{s}_{N-1}\rangle,
\end{equation}
which is illustrated in Fig.~\ref{fig:tJ_TN}(b).
$N$ is the number of sites.
$T$ and $\Lambda$ are all fermionic tensors with bond dimension $D$.
The corresponding tensor trace "$f\text{-Tr}$" should follow the canonical fermionic rule~\cite{PhysRevB.88.115139, PhysRevB.95.075108, Bultinck2018}, contracting all internal virtual bonds in the network.
$\{s\}$ denotes a physical configuration with each local Hilbert space being spanned by three orthogonal states:
vacuum $\ket{0}$, $\ket{\uparrow}= c_{\uparrow}^{\dagger}\ket{0}$, and $\ket{\downarrow}=c_{\downarrow}^{\dagger}\ket{0}$.
Doping is controlled by the chemical potential $\mu$ in a grand canonical ensemble.
\begin{figure}[t]
    \centering
    \includegraphics[width=0.48\textwidth]{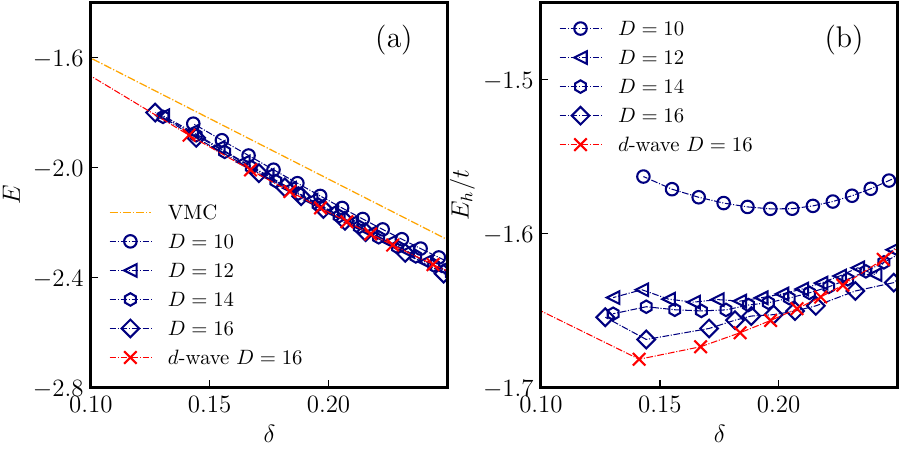}
    \caption{
    Energy comparison of $d$-wave (red data) and $s$-wave (blue data) states in the square $t$-$J$ model with $t/J=3.0$.
    (a) Per-site energy $E(\delta)$.
    The dashed line represents previous variational Monte Carlo (VMC) results~\cite{PhysRevB.70.104503, PhysRevB.81.165104}.
    (b) Per-hole energy $E_h(\delta)\equiv\left[E(\delta)-E_0\right]/\delta$.
    $E_{0}$ is the energy of half-filled Heisenberg model~\cite{PhysRevB.56.11678}.
    }
    \label{fig:tJ_swave_ene}
\end{figure}

\begin{figure}[t]
    \centering
    \includegraphics[width=0.32\textwidth]{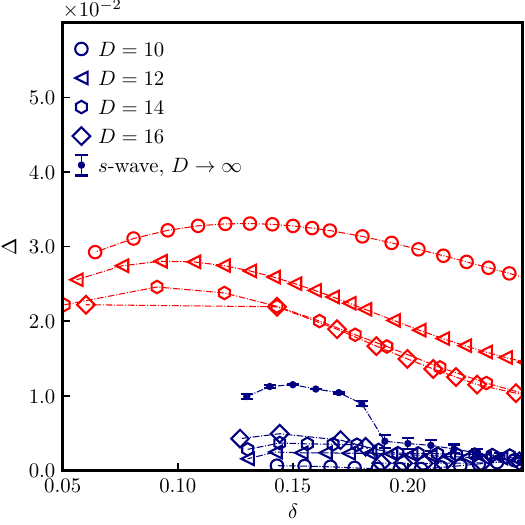}
    \caption{
    Superconducting Cooper pair amplitude $\Delta$ of the $d$-wave (red data) and $s$-wave (blue data) states in the square $t$-$J$ model with $t/J=3.0$.
    $\Delta$ of the $d$-wave state almost saturates up to $D=16$, thus we regard the data from $D=16$ as $D\rightarrow\infty$ for the $d$-wave.
    Estimated $\Delta$ of the $s$-wave state in $D\rightarrow\infty$ are linearly extrapolated with $1/D$.
    }
    \label{fig:tJ_swave_sc}
\end{figure}

\paragraph*{The emergence of s-wave paring superconductivity at large doping --- }
In the main text, we focus on $t/J=3.0$ as experiments suggested~\cite{RevModPhys.78.17}.
Starting from many randomly initialized states $\ket{\Psi}$, we employ the imaginary-time evolution method to approach a ground state $\ket{\Psi_{0}} \propto\lim_{\tau\rightarrow\infty}e^{-\tau{H}}\ket{\Psi}$.
Fermionic cluster update methods~\cite{PhysRevLett.101.090603, 2411.19218} are utilized.
By employing this improved infinite fermionic tensor network method, in addition to the well-known $d$-wave state~\cite{PhysRevB.61.R11894, PhysRevLett.127.097003, PhysRevLett.113.046402}, recently we discovered several other new pair density wave (PDW) states~\cite{2411.19218} in the underdoped regime of the $t$-$J$ model, indicating that the ground states subspace of this seemingly simple model could be highly degenerate and much more complicated than previously anticipated.
At larger doping, unexpectedly, we obtain the $s$-wave states as one of the local minimums alongside with other competing orders~\cite{PhysRevLett.113.046402, 2411.19218}.

As shown in Fig.~\ref{fig:tJ_swave_ene}, energies of the $s$-wave states are comparable with that of $d$-wave states at $\delta\gtrsim 15\%$ up to $D=16$ as the limit of our computational power.
In Fig.~\ref{fig:tJ_swave_sc}, for the superconducting singlet pair amplitude $\Delta$, interestingly, we find that although the $d$-wave pairing amplitude $\Delta_{d}$ could be much larger than $s$-wave pairing $\Delta_{s}$ with the same bond dimension $D$, their dependence $D$ is opposite.
That is, with increasing $D$, the amplitude of $\Delta_{s}$ increases, as shown in Fig.~\ref{fig:tJ_swave_sc}, while $\Delta_{d}$ decreases and almost saturates for $D=16$~\cite{2411.19218}.
We believe that the $s$-wave states are most likely to survive as a competing order in the overdoped regime $15\%\sim 25\%$ for the $t$-$J$ model. Of course, the bare $t$-$J$ model is not sufficient to understand realistic systems, and we will investigate the effect of other perturbations in the following. 


\begin{figure}[tbh]
    \centering
    \includegraphics[width=0.25\textwidth]{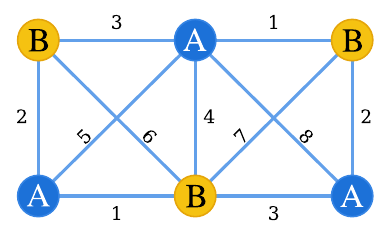}
    \includegraphics[width=0.45\textwidth]{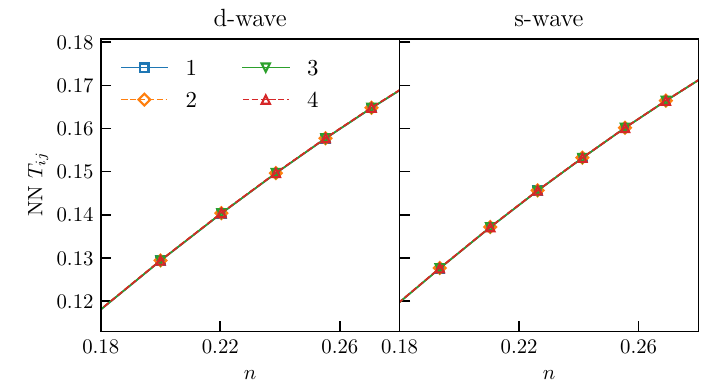}
    \includegraphics[width=0.45\textwidth]{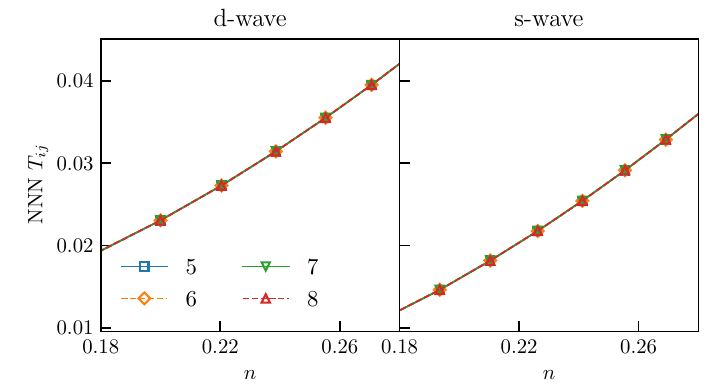}
    \caption{
    The hopping $T_{ij} = \sum_\sigma \braket{c^\dagger_{i\sigma} c_{i\sigma}}$ on NN and NNN bonds in the $d$-wave and the $s$-wave states with $D = 12$ in doping range $0.18 \le n \le  0.28$, measured with CTM boundary dimension $\chi = 64$. 
    }
    \label{fig:nnn-hopping}
\end{figure}

\paragraph*{The effect of NNN hopping and three-site term --- }
It has been shown that the next-nearest neighbor (NNN) hoping plays a very important role for the asymmetry of hole and electron doping in materials, and it becomes more and more important at large doping. 
To measure NNN bonds, it is more convenient to employ the corner transfer matrix renormalization group algorithm (CTMRG) \cite{Corboz2014}. 
Our analysis reveals an intriguing result: the NNN hopping energy associated with the s-wave state is slightly lower than that of the conventional d-wave state (Fig. \ref{fig:nnn-hopping}). This observation suggests that 
a surface reconstruction featuring larger (negative) $t^\prime$ has the potential to stabilize the s-wave state, allowing for its microscopic coexistence with the d-wave state at large doping. 

\begin{figure}[tb]
    \centering
    \includegraphics[width=0.48\textwidth]{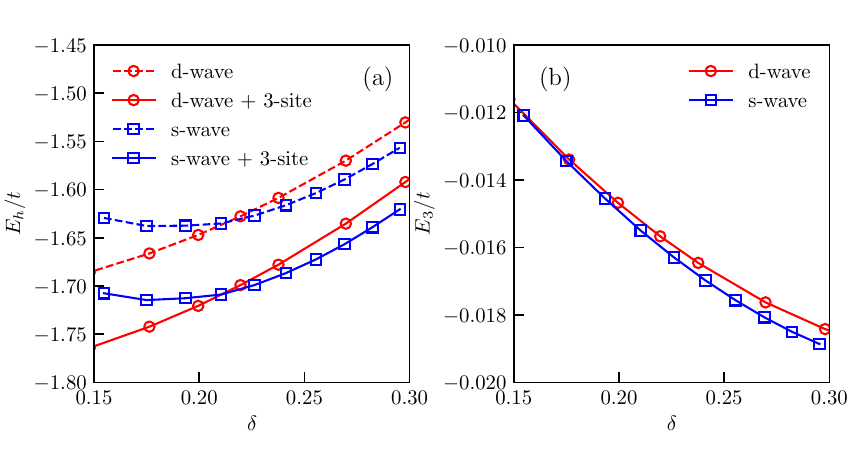}
    \caption{
    The three-site energy $E_3 = \braket{H_3}$ with $\alpha = J/4$ for the $d$-wave and the $s$-wave states with $t/J = 3$ and $D = 12$, measured with CTM boundary dimension $\chi = 64$. (a) Energy per hole without and with the three-site term. (b) The three-site energy per site. 
    }
    \label{fig:3site}
\end{figure}

In addition to the $t$-$J$ terms, the large-$U$ limit of the Hubbard model also contains the so-called three-site hopping term \cite{Ogata2008}
\begin{equation}
    H_3 = -\alpha 
    \sum_{\substack{
        \braket{i,j,k}, i \ne k , \sigma
    }} P_G (
        c^\dagger_{i\sigma} c^\dagger_{j\bar{\sigma}}
        c_{j\bar{\sigma}} c_{k\sigma}
        - c^\dagger_{i\sigma} c^\dagger_{j\bar{\sigma}}
        c_{j\sigma} c_{k\bar{\sigma}}
    ) P_G,
\end{equation}
where $\alpha = J/4$, and $\braket{i,j,k}$ indicates that $i,k$ are the nearest neighbor of $j$. Mean field theory \cite{Ruckenstein1987,Jedrak2011,Yang2024}, VMC \cite{Li1993} and DMRG \cite{Yang2024} calculations have shown that in the overdoped region, the three-site term suppresses the $d$-wave SC and enhances the $s$-wave pairing instead. 
Fig. \ref{fig:3site} shows that the three-site energy $E_3 = \braket{H_3}$ is lower at large doping $\delta \gtrsim 0.15$ in the $s$-wave state than in the $d$-wave state. 
In fact, if we start from the three-band Hubbard model, we will end up with a much larger $\alpha$ which can further stabilize the $s$-wave state.

\paragraph*{Ginzburg-Landau Theory --- } In the following, we study the
interplay between the $d$-wave and $s$-wave superconductivity in the twisted bilayer cuprates with the Ginzburg-Landau theory based on the above calculations.
The free energy density is given by
$
    f = f_0(\psi^t_s,\psi^t_d)
    + f_0(\psi^b_s,\psi^b_d) 
    + f_\text{int}\left(\psi_s^{t/b},\psi_d^{t/b}\right),
    \label{Eq:fe}
$
where $\psi^{t(b)}_{s(d)}$ is the
$s(d)$-wave superconducting order parameter of the top(bottom) layer.
$f_0$ is the free energy density of each single layer.  To include both the $s$-wave and the $d$-wave pairing channels, we take the following form for $f_0$:
\begin{align}
    f_0(\psi_s,\psi_{d}) =&-\alpha_s |\psi_s|^2-\alpha_d |\psi_{d}|^2+ \frac{\beta_s}{2}|\psi_s|^4+ \frac{\beta_d}{2}|\psi_{d}|^4 \nonumber\\
    &+ \beta_{sd}|\psi_s|^2|\psi_d|^2 ,
\end{align}
where $\alpha$ and 
$\beta$ are Ginzburg-Landau coefficients.  Since no coexistence of $s$-wave and $d$-wave order is found in our tensor network calculations, we choose $\beta_{sd} = \sqrt{\beta_s \beta_d}$ such that the system favors either a pure $s$-wave (when $\alpha_s^2/\beta_s>\alpha_d^2/\beta_d$) or a pure $d$-wave state (when $\alpha_d^2/\beta_d>\alpha_s^2/\beta_s$).


The Josephson coupling between the twisted bilayers are encoded in $f_{\text{int}}$:
\begin{equation}\label{Int.term}
f_\text{int}\left(\psi_s^{t/b},\psi_d^{t/b}\right)= -J \psi_s^{t*}\psi_s^{b} - J \cos(2\theta) \psi_d^{t*}\psi_d^{b} + h.c.,
\end{equation}
where $J > 0$ is the Josephson coupling strength for the $s(d)$-wave superconducting order parameter, and $\theta$ is the twist angle. The layer-exchange symmetry $(t \leftrightarrow b, \theta \rightarrow -\theta)$ leads to $|\psi_{d}^t|=|\psi_{d}^b| \equiv \psi_d$, $|\psi_{s}^t|=|\psi_{s}^b|\equiv \psi_s$. For simplicity, we take $\psi_{d}^{t/b}=\psi_{d} e^{\pm i\phi_d/2}$ and $\psi_{s}^{t/b}=\psi_{s} e^{\pm i\phi_s / 2}$, where $\phi_s$ ($\phi_d$) is the phase difference between the top and bottom layers for $s$-wave (d-wave) superconducting order parameter. The total free energy density reads:
\begin{align}
    f=&-\left(2\alpha_s+2J\cos\phi_s\right)\psi_s^2-\left(2\alpha_d+2J\cos (2\theta)\cos \phi_d \right)\psi_{d}^2\nonumber \\ +& \left(\sqrt{\beta_s}\psi_s^2+\sqrt{\beta_d}\psi_{d}^2\right)^2.
\end{align}
The free energy density is minimized at $\phi_s = 0$ with $\phi_d = 0$ when $\cos (2\theta) > 0$, but shifts to $\phi_d = \pi$ when $\cos (2\theta) < 0$, where the system exhibits $s$-wave superconductivity if $(\alpha_d + J|\cos (2\theta)|)^2 / \beta_d < (\alpha_s + J)^2 / \beta_s$, and $d$-wave superconductivity if $(\alpha_d + J|\cos (2\theta)|)^2 / \beta_d > (\alpha_s + J)^2 / \beta_s$.

In the case where the energy of the $s$-wave and $d$-wave is close, i.e., $(\alpha_d+J)^2/\beta_d \gtrsim (\alpha_s+J)^2/\beta_s$, there will be a transition between the $d$-wave and $s$-wave at the twist angle:
\begin{equation}\label{GL:criticalangle}
\theta_c=\frac{1}{2}\arccos\left(\frac{\psi_s^2}{\psi_d^2}\right).
\end{equation}
Here, $\psi_{s/d} = \sqrt{\alpha_{s/d}/\beta_{s/d}}$ and we assumed that $f_d=f_s$. Since $J$ is contracted, $\theta_c$ is independent of $J$. The detailed derivation is in the Supplementary Material.
We assume that the superconducting order parameter $\psi_{s/d}$ is proportional to the corresponding superconducting Cooper pair amplitude of the tensor network calculations. Here, we select a doping concentration of $\delta = 0.15$ as representative of the optimally doped regime, and $\delta = 0.18$ for the overdoped regime, as these values are close to the experimental data and well defined under $1/D$ scaling.



\paragraph*{Josephson Current Density --- }\label{para:Josephsoncurrent}
To compute the Josephson current density, we consider the case with finite $\phi_s$ or $\phi_d$, and the free energy density can be written as~\cite{JosephonCurrent1}
\begin{equation}
f(\phi_s,\phi_d)=E_0-2 J\left(\psi_s^2 \cos\phi_s+ \cos (2\theta)\psi_d^2 \cos\phi_d\right),
\end{equation}
where $E_0$ contains the terms that are independent of $\phi_{s/d}$. The Josephson current density for $s$($d$)-wave pairing is given by
\begin{equation}\label{GL_current}
j_{s/d}(\phi_{s/d})=\frac{2e}{\hbar}\frac{\partial f}{\partial\phi_{s/d}} = j_c^{s/d} \sin \phi_{s/d},
\end{equation}
where the $j_c^{s} =4eJ\psi_s^2/\hbar$ and $j_c^{d} =4eJ\psi_d^2|\cos(2\theta)|/\hbar$.
In the presence of a transition from d-wave to s-wave at twist angle $\theta_c$, the critical current density at a certain twist angle $\theta$ is $j_c^s$ for $\theta_c<\theta<\frac{\pi}{2}-\theta_c$ and $j_c^d$ otherwise.

\begin{figure}[tb]
\centering\includegraphics[width=0.48\textwidth]{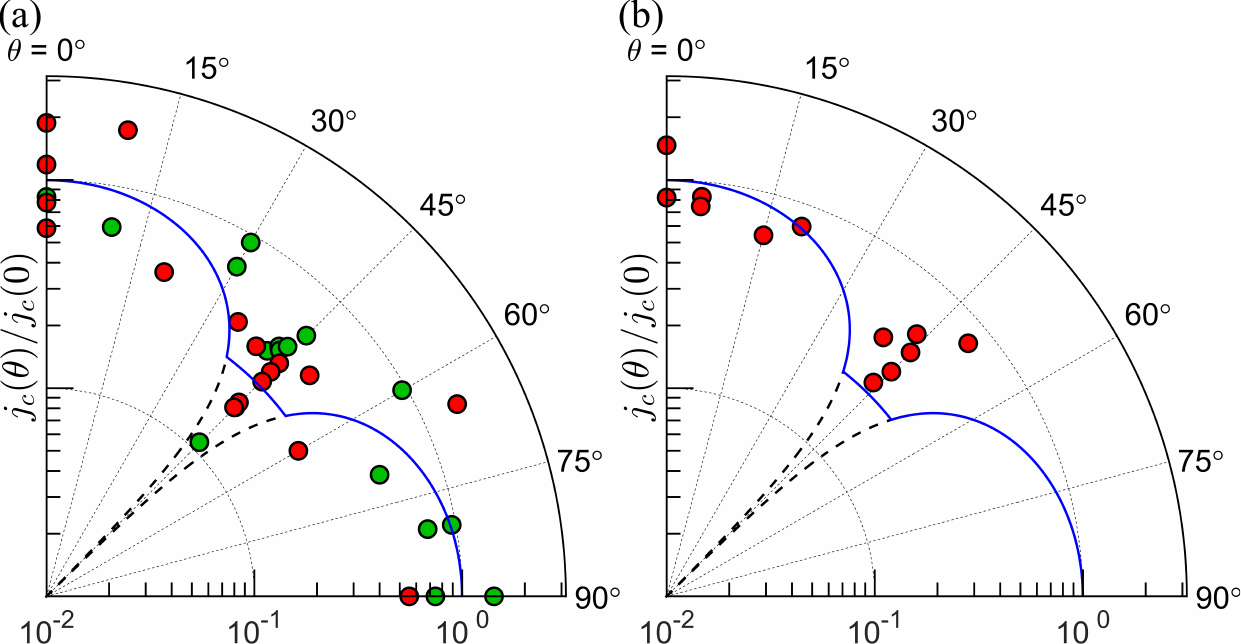}
    \caption{Critical Josephson current density $j_c$ as a function of twist angle $\theta$ for (a) the optimally doped and (b) overdoped regimes. The blue solid and dashed lines represent the results from Ginzburg-Landau theory and the pure $d$-wave state, respectively. The experimental data are taken from the following references: the red points from Ref.~\cite{PhysRevB.108.174508} and the green points from Ref.~\cite{Wang2023}.}
    \label{fig:GL.current}
\end{figure}

The critical Josephson current density $j_c$ vs twist angle $\theta$ is depicted in Fig. \ref{fig:GL.current}.  $j_c(\theta)$ exhibits nonsmoothness at $\theta=\theta_c$ and $\theta=\pi/2-\theta_c$, corresponding to a transition between $d$-wave and $s$-wave pairing. A comparison between our Ginzburg-Landau theory calculations and the experimental data from Refs.~\cite{PhysRevB.108.174508,Wang2023} is presented. To maintain consistency with the experimental data format, a logarithmic scale was used in the polar plots. Qualitative agreement is observed between our result and experiment: the $\delta = 0.15$ case (a, blue solid line) corresponds to the optimally doped regime, while the $\delta = 0.18$ case (b, blue solid line) corresponds to the overdoped regime.

\begin{figure}[tb]
    \centering
  \includegraphics[width=0.3\textwidth]{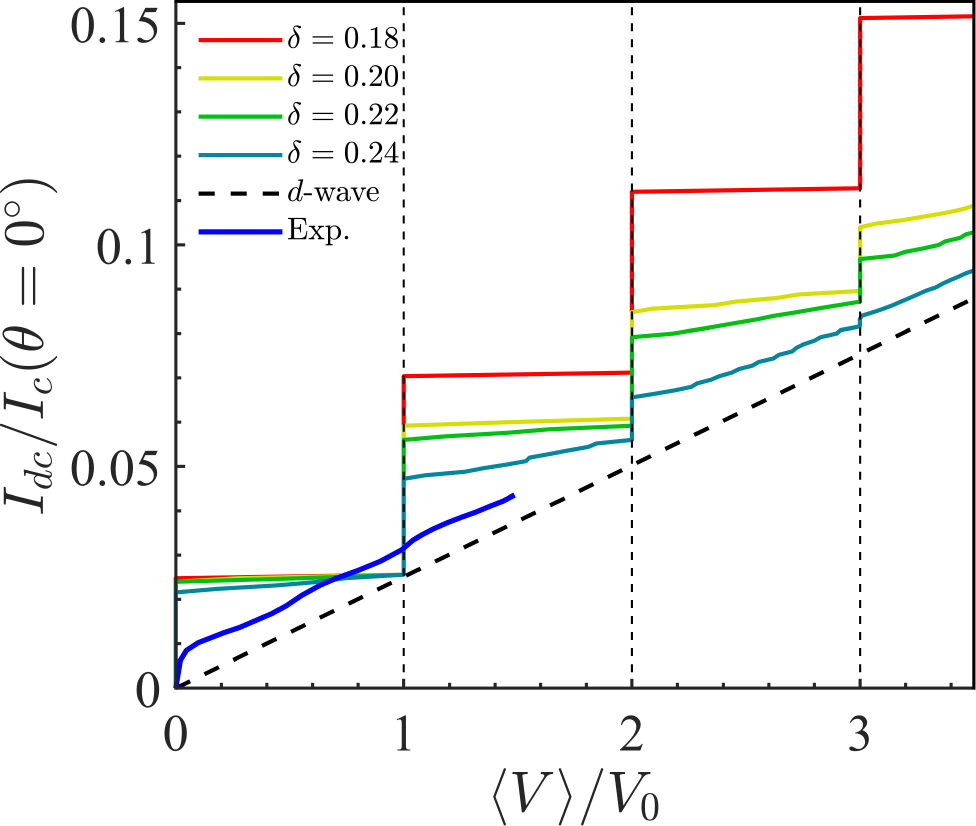}
    \caption{Numerical simulation results of Shapiro steps behavior in the overdoped regime at twist angle $\theta=45^\circ$. The experimental data are taken from Ref.~\cite{science.abl8371}. The system parameters are chosen to match the experimental conditions and to generate well-defined Shapiro steps: $I_{ac}=I_c(\theta=45^\circ)$, $\omega_m/2\pi = 40$ GHz, $R = 20.68$ $\Omega$.}
    \label{fig:GL.Shapiro}
\end{figure}

\paragraph*{Shapiro steps --- }
Finally, we discuss the AC Josephson effect with the so-called resistively shunted junction (RSJ) model~\cite{RSCJ1,RSCJ2}
\begin{equation}
     I(t)=I_c \sin{\phi}+\frac{\Phi_0}{2\pi R} \frac{d\phi}{dt},
\end{equation}
where the external drive takes the form $I(t)=I_{dc}+I_{ac} \sin(\omega_{m} t)$ and $I=jS$. Here, $S$ is the area of the junction, $R$ is the junction resistance, $\omega_{m}$ is the frequency of the microwave,  and $\Phi_0=h/2e$ is the flux quantum. The characteristic curve exhibits integer Shapiro steps~\cite{shapirostep} at voltage intervals $\langle V_n \rangle=n V_0~~~(n\in \mathbb{Z})$, where $V_0\equiv \Phi_0 \frac{\omega_m}{2\pi}$. Since we neglect the higher-order terms in Eq. \eqref{Int.term}, our analysis is restricted to the integer steps. At a twist angle of $45^\circ$, the steps are expected to be suppressed for the $d$-wave case, while they should remain present for the $s$-wave case. Here, the persistence of the steps at $\theta=45^\circ$ is attributed to the $d$-wave to $s$-wave transition. The numerical simulation result is shown in Fig. \ref{fig:GL.Shapiro}. The observability of the integer Shapiro steps depends on the magnitude of the critical current density at $\theta=45^\circ$, that is $I_c(45^\circ)$. Our results focus on the overdoped regime, where robust integer steps are observed due to a predominant $s$-wave pairing propensity. In contrast, the optimally doped regime exhibits strong $d$-wave pairing, which leads to poorly defined integer steps. Although experimental data from Ref.~\cite{science.abl8371} show no Shapiro steps, which may be attributed to the small $I_c(45^\circ)$. The experimentally observed Fiske steps \cite{Wang2023} provide independent evidence for the emergence of $s$-wave pairing, consistent with our results of integer steps. 

\paragraph*{Conclusion and discussion --- }
In this paper, we performed the state-of-the-art tensor network simulation for overdoped $t$-$J$ model. To our surprise, 
we find that both $d$-wave and $s$-wave pairing symmetries are possible for overdoped cuprates. 
In particular, the variational energy of the $s$-wave state becomes comparable to the $d$-wave state when $\delta\gtrsim 15\%$. Moreover, the negative NNN hopping term and the so-called three-cite terms will further stabilize the $s$-wave state.
Although with the same bond dimension $D$, the pairing magnitude of the $d$-wave state is much larger than that of the $s$-wave, their dependence trends $D$ are opposite, which indicates that the $s$-wave pairing will be finite and close to the $d$-wave as $D\rightarrow\infty$. As a simple application, we further developed a Ginzburg-Landau theory with almost degenerate $s$-wave and $d$-wave pairing free energy to compute the Josephson current density measured for twisted bilayer cuprate systems. Our results qualitatively explain the experimentally observed deviation from the pure $d$-wave pairing state. We further compute the Shapiro steps at different doping concentrations with fixed twisted angle $\theta=45^\circ$, which can be carefully examined by future experiments. 

\paragraph*{Acknowledgments --- }
WZ acknowledges discussions with Yi Tan, Jia-Wei Mei and Shuo-Ying Yang.  
The CTMRG measurement is performed using the \texttt{PEPSKit.jl} package \cite{pepskit}. This work is supported by funding from Hong Kong's Research Grants Council (CRF C7012-21GF and GRF No. 14302021), the National Key Research and Development Program of China (No. 2024YFA1408101), NSFC (Grants No. 12141402, 12334002, 12504176), Guangdong Provincial Quantum Science Strategic Initiative Grand No. SZZX2401001, the SUSTech-NUS Joint Research Program, the Science, Technology and Innovation Commission of Shenzhen Municipality (No. ZDSYS20190902092905285), and Center for Computational Science and Engineering at Southern University of Science and Technology.

\appendix

\section{Supplementary data from fermionic tensor network simulations}

\begin{figure}[h]
    \centering
    \includegraphics[width=0.48\textwidth]{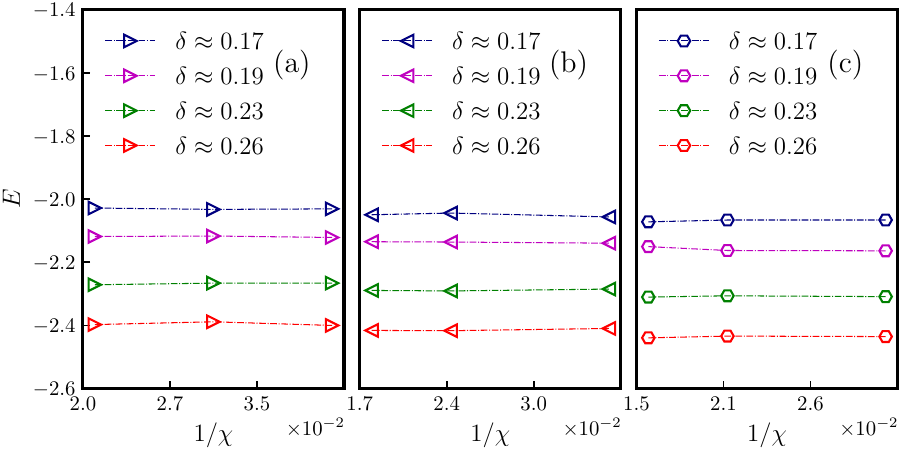}
    \caption{
    Boundary MPS convergence benchmark for some $s$-wave states with different doping level $\delta$ in the square $t$-$J$ model with $t/J=3.0$.
    $\chi$ is the bond dimension of the boundary MPS.
    (a) $D=12$.
    (b) $D=14$.
    (c) $D=16$.
    }
    \label{fig:tJ_vumps_convergence}
\end{figure}

Firstly, we perform some benchmark to make sure our VUMPS measurement is reliable.
As shown in Fig.~\ref{fig:tJ_vumps_convergence}, with $t/J=3.0$, for some typical $s$-wave states at large doping levels around $17\%$ to $26\%$, different VUMPS bond dimensions $\chi$ give relatively well converged energies.
VUMPS needs smaller $\chi$ than CTMRG because it is a variational method rather than projections.

\begin{figure}[tbh]
    \centering
    \includegraphics[width=0.48\textwidth]{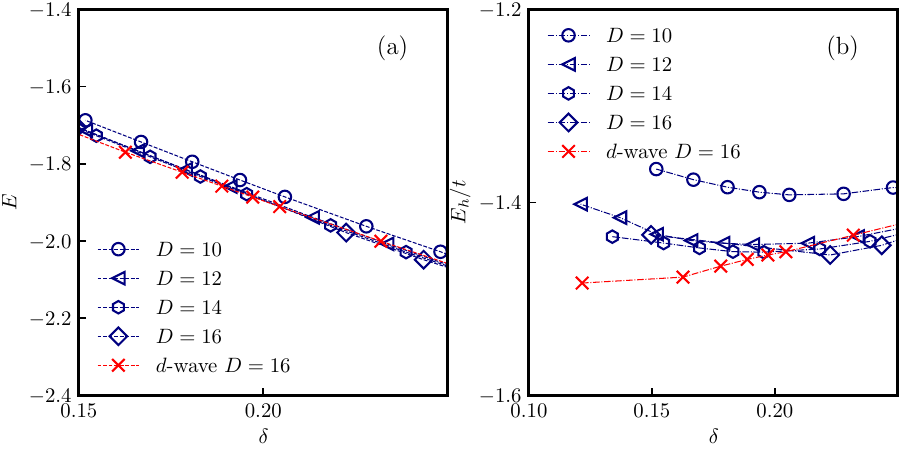}
    \caption{
    Energy comparison of $d$-wave (red data) and $s$-wave (blue data) states in the square $t$-$J$ model with $t/J=2.5$.
    (a) Per-site energy $E(\delta)$.
    (b) Per-hole energy $E_h(\delta)\equiv\left[E(\delta)-E_0\right]/\delta$.
    $E_{0}$ is the energy of half-filled Heisenberg model~\cite{PhysRevB.56.11678}.
    }
    \label{fig:tJ_swave_ene_t25}
\end{figure}

\begin{figure}[tbh]
    \centering
    \includegraphics[width=0.3\textwidth]{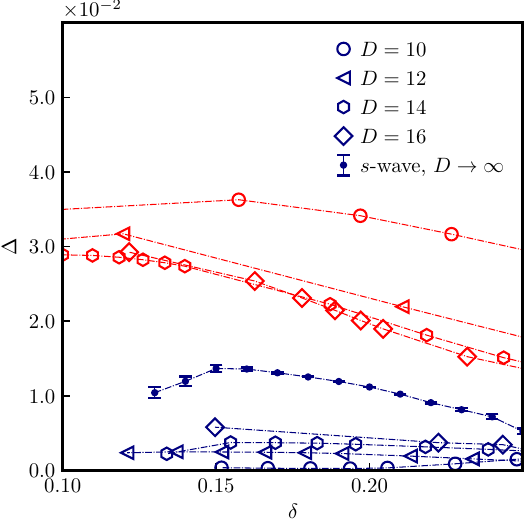}
    \caption{
    Superconducting Cooper pair amplitude $\Delta$ of the $d$-wave (red data) and $s$-wave (blue data) states in the square $t$-$J$ model with $t/J=2.5$.
    $\Delta$ of the $d$-wave state almost saturates up to $D=16$, thus we regard the data from $D=16$ as $D\rightarrow\infty$ for the $d$-wave.
    Estimated $\Delta$ of the $s$-wave state in $D\rightarrow\infty$ are linearly extrapolated with $1/D$. 
    }
    \label{fig:tJ_swave_sc_t25}
\end{figure}

Next, we provide more data for $t/J=2.5$, as shown in Figs.~\ref{fig:tJ_swave_ene_t25} and~\ref{fig:tJ_swave_sc_t25}.
In Fig.~\ref{fig:tJ_swave_ene_t25}, energies of the $s$-wave states with $t/J=2.5$ are comparable with that of $d$-wave states at $\delta\gtrsim 17\%$ up to $D=16$, which is slightly different with $t/J=3.0$.
It looks like that $s$-wave state could survive better than the $d$-wave state with a larger $t/J$ ratio.
In Fig.~\ref{fig:tJ_swave_sc_t25}, for the superconducting singlet pair amplitude $\Delta$, we find a similar conclusion with $t/J=3.0$.
The $d$-wave pairing amplitude $\Delta_{d}$ is much larger than $s$-wave pairing $\Delta_{s}$ with the same $D$, but their dependence $D$ is opposite.
With increasing $D$, the amplitude of $\Delta_{s}$ increases, while $\Delta_{d}$ decreases and almost saturates for $D=16$.

\section{Detailed Derivation of the Ginzburg-Landau Theory}
In the following, we provide the details regarding the Ginzburg–Landau theory, which primarily consists of two parts: first, determining the locations and values of the minima of the free energy density; second, identifying the critical twist angle.
 
The free energy density of the Ginzburg-Landau theory for a single layer is given by
\begin{equation}\label{eq:slfd}
f_0(\psi_s,\psi_{d})=-\alpha_s \psi_s^2-\alpha_d \psi_d^2+ \big(\sqrt{\frac{\beta_s}{2}}\psi_s^2+\sqrt{\frac{\beta_d}{2}}\psi_d^2\big)^2.
\end{equation}
The locations of the local minima are determined by taking the partial derivative of the free energy density with respect to the order parameter $\psi_s$ and setting the result to zero. That is
\begin{equation}
	\frac{\partial f_0}{\partial \psi_s}=-2\alpha_s \psi_s+4\big(\sqrt{\frac{\beta_s}{2}}\psi_s^2+\sqrt{\frac{\beta_d}{2}}\psi_d^2\big)\sqrt{\frac{\beta_s}{2}}\psi_s=0.
\end{equation}
Then, the locations of minima are
\begin{equation}
	\psi_s=0,
\end{equation}
and
\begin{equation}
	\psi_s=\pm \sqrt{\frac{\alpha_s-\sqrt{\beta_s\beta_d}\psi_d^2}{\beta_s}}.
\end{equation}
Similarly, in the case of $\psi_d$, the locations of the local minima are given by
\begin{equation}
	\psi_d=0,
\end{equation}
and
\begin{equation}
	\psi_d=\pm \sqrt{\frac{\alpha_d-\sqrt{\beta_d\beta_s}\psi_s^2}{\beta_d}}.
\end{equation}
For $\psi_s=0$, $\psi_d=\pm\sqrt{\frac{\alpha_d}{\beta_d}}$, the value of the local minima is $f_{\text{min}}=-\frac{\alpha_d^2}{2\beta_d}$, while $\psi_d=0$, $\psi_s=\pm\sqrt{\frac{\alpha_s}{\beta_s}}$, the value of the local minima is $f_{\text{min}}=-\frac{\alpha_s^2}{2\beta_s}$. For $\psi_s=\pm \sqrt{\frac{\alpha_s-\sqrt{\beta_s\beta_d}\psi_d^2}{\beta_s}}$ and $\psi_d=\pm \sqrt{\frac{\alpha_d-\sqrt{\beta_d\beta_s}\psi_s^2}{\beta_d}}$, these minima exist only when $\alpha_d^2/\beta_d=\alpha_s^2/\beta_s$ owing to the singular matrix. It means that if $\alpha_d^2/\beta_d=\alpha_s^2/\beta_s$, the locations of the minima become $\psi_s^2\sqrt{\beta_s}+\psi_d^2\sqrt{\beta_d}=\alpha_s/\sqrt{\beta_s}$, which is the elliptic equation. This result is consistent with tensor network calculations, which indicate that the $s$-wave and $d$-wave superconducting states are accidentally degenerate. Hence, the locations and value of the global minimum are governed by the magnitudes of $-\frac{\alpha_s^2}{2\beta_s}$ and $-\frac{\alpha_d^2}{2\beta_d}$. Notably, the conclusion for the minima of the free energy density remains unchanged even upon adding a term $c \psi_s^2 \psi_d^2$ with an arbitrarily large positive coefficient $c$. This robustness signifies that the model inherently describes a strong competition between the $s$-wave and $d$-wave pairing channels. To prove this, Eq. \eqref{eq:slfd} becomes
\begin{equation}
        f_0(u,v)=-\alpha_s u-\alpha_d v+\frac{\beta_s}{2}u^2+\frac{\beta_d}{2}v^2+cuv.
\end{equation}
Here, $u\equiv \psi_s^2$, $v\equiv \psi_d^2$, $K\equiv \sqrt{\beta_s\beta_d}+c$. Since
\begin{equation}
    \begin{split}
        &\frac{\partial f_0}{\partial u}=-\alpha_s+Kv+\beta_s u=0,\\&
        \frac{\partial f_0}{\partial v}=-\alpha_d+Ku+\beta_d v=0,
    \end{split}
\end{equation}
the location of the local minimum is given by
\begin{equation}
    \begin{split}
        &u_m=\frac{\alpha_s \beta_d-K \alpha_d}{\beta_s \beta_d-K^2},\\&v_m=\frac{\alpha_d \beta_s-K \alpha_s}{\beta_s \beta_d-K^2}.
    \end{split}
\end{equation}
Using the relation
\begin{equation}
    \begin{split}
        &\alpha_s=\beta_s u_m+K v_m,\\&\alpha_d=K u_m+\beta_d v_m,
    \end{split}
\end{equation}
the value of the local minimum $f_m$ is given by
\begin{equation}
    \begin{split}
        f_m&=-(\beta_s u_m+K v_m)u_m-(K u_m+\beta_d v_m)v_m\\&+Ku_mv_m+\frac{\beta_s}{2}u_m^2+\frac{\beta_d}{2}v_m^2\\&=-\left(Ku_mv_m+\frac{\beta_s}{2}u_m^2+\frac{\beta_d}{2}v_m^2\right).
    \end{split}
\end{equation}
Finally, we compare $f_m$ with $f_d$ and $f_s$. That is
\begin{equation}
    \begin{split}
        f_m-f_d&=-\left(Ku_mv_m+\frac{\beta_s}{2}u_m^2+\frac{\beta_d}{2}v_m^2\right)-\left(-\frac{\alpha_d^2}{2\beta_d}\right)\\&=-\left(Ku_mv_m+\frac{\beta_s}{2}u_m^2+\frac{\beta_d}{2}v_m^2\right)+\frac{(K u_m+\beta_d v_m)^2}{2\beta_d}\\&=\frac{u_m^2}{2\beta_d}(K^2-\beta_d\beta_s)> 0.
    \end{split}
\end{equation}
Here, $K^2-\beta_d\beta_s=c^2+2c\beta_d\beta_s>0$.
Similarly,
\begin{equation}
    f_m-f_s=\frac{v_m^2}{2\beta_s}(K^2-\beta_d\beta_s)> 0.
\end{equation}
Therefore, $f_m>\text{min}[f_d,f_s]$, $f_m$ is not the minimal value when $c>0$.

\section{The critical twist angle}
To calculate the critical twist angle, we need to define the difference between $f_s$ and $f_d$, i.e., $\Delta f\equiv f_{d}-f_{s}$, where $f_{s/d}\equiv -\alpha_{s/d}^2/2\beta_{s/d}$. Therefore, the ratios $\alpha_s/\alpha_d$ and $\beta_d/\beta_s$ can be expressed in terms of the ratios $\Delta f/|f_d|$ and $\psi_d^2/\psi_s^2$. They are given by
\begin{equation}\label{SMGL1}
    \begin{split}
       \frac{\alpha_{d}^2}{\beta_{d}}&=\frac{\alpha_{s}^2}{\beta_{s}}-2\Delta f\\
\frac{\alpha_{d}^2}{\alpha_{s}^2}&=\frac{\beta_{d}}{\beta_{s}}- \frac{\alpha_d^2}{\alpha_{s}^2}\frac{\Delta f}{|f_d|}\\
\frac{\beta_{d}}{\beta_{s}}=&\left(1+\frac{\Delta f}{|f_d|}\right)\frac{\alpha_{d}^2}{\alpha_{s}^2},
    \end{split}
\end{equation}
and
\begin{equation}\label{SMGL2}
    \begin{split}
       \frac{\alpha_{d}^2}{\beta_{d}}&=\frac{\alpha_{s}^2}{\beta_{s}}-2\Delta f\\
\psi_d^2\alpha_d&=\psi_s^2\alpha_s-2\Delta f\\
\frac{\alpha_s}{\alpha_d}&=\left(1+\frac{\Delta f}{|f_d|}\right)\frac{\psi_d^2}{\psi_s^2}.
    \end{split}
\end{equation}
The critical twist angle can be determined as the angle at which the minima of the free energy density of the $s$-wave and $d$-wave states become equal. That is
\begin{equation}\label{SMGL3}
    \begin{split}
        \frac{(\alpha_s+J)^2}{2\beta_s}&=\frac{(\alpha_d+J|\cos (2\theta_c)|)^2}{2\beta_d}\\
|\cos(2\theta_c)|&=\frac{\sqrt{\frac{\beta_d}{\beta_s}}(\alpha_s+J)-\alpha_d}{J}\\
\theta_c&=\frac{1}{2}\arccos\left(\frac{\sqrt{\frac{\beta_d}{\beta_s}}(\alpha_s+J)-\alpha_d}{J}\right).
    \end{split}
\end{equation}
Here, $J$ is the Josephson coupling strength. $\theta_c$ is, by definition, bounded between $0$ and $\pi/4$; consequently, any value outside this interval is physically meaningless as it would not induce a transition. After we substitute Eqs. \eqref{SMGL1}, \eqref{SMGL2} into Eq. \eqref{SMGL3}, the critical twist angle $\theta_c$ becomes

\begin{equation}\label{thetac}
    \begin{split}
    \theta_c&=\frac{1}{2}\arccos\left(\frac{\sqrt{\frac{\beta_d}{\beta_s}}(\alpha_s+J)-\alpha_d}{J}\right)\\
&=\frac{1}{2}\arccos\left(\frac{\sqrt{\left(1+\frac{\Delta f}{|f_d|}\right)}\left[1+\left(\left(1+\frac{\Delta f}{|f_d|}\right)^{-1}\frac{\psi_s^2}{\psi_d^2}\right)J/\alpha_d\right]-1}{J/\alpha_d}\right)\\
&=\frac{1}{2}\arccos\left(\left(\sqrt{1+\frac{\Delta f}{|f_d|}}-1\right)\frac{1}{J/\alpha_d}+\frac{\psi_s^2}{\psi_d^2}\left(1+\frac{\Delta f}{|f_d|}\right)^{-\frac{1}{2}}\right).
    \end{split}
\end{equation}

It is clear that there is no transition in the limit of weak Josephson coupling ($J\rightarrow 0$) and the system retains its intrinsic pairing symmetry depending on the sign of $\Delta f$. In contrast, in the limit of strong Josephson coupling ($J\rightarrow +\infty$), one has $\theta_c\rightarrow \arccos\left((\psi_s^2/\psi_d^2)/\sqrt{1+\Delta f/|f_d|}\right)/2$. The transition exists when $\Delta f>f_d$ alongside constraint $\psi_s^2/\psi_d^2\le \sqrt{1+\Delta f/|f_d|}$. If we take $\Delta f=0$, Eq. \eqref{thetac} will become Eq. \eqref{GL:criticalangle}. It means that $\theta_c$ can be represented by the critical Josephson current density $j_c(\theta)$, that is $\theta_c=\arccos(j_c(45^\circ)/j_c(0^\circ))/2$.

\clearpage
\bibliographystyle{apsrev4-2} 
\bibliography{refs}

\end{document}